\documentclass[twocolumn,english,showpacs,preprintnumbers,superscriptaddress,prb]{revtex4-1}
\usepackage[english]{babel}
\usepackage{setspace}
\usepackage{graphicx}
\usepackage{amssymb}
\usepackage{amscd}
\usepackage{rotating}
\usepackage[usenames, dvipsnames]{color}
\usepackage{epstopdf}
\usepackage{comment}
\usepackage{siunitx}
\usepackage{float}
\usepackage{xfrac}
\usepackage{natbib}
\usepackage{hyperref}
\hypersetup{
    colorlinks=true,
    linkcolor=blue,
    filecolor=blue,
    urlcolor=blue,
   citecolor=blue,
}
\usepackage{color}
\usepackage[normalem]{ulem}
\newcommand\footnoteref[1]{\protected@xdef\@thefnmark{\ref{#1}}\@footnotemark}

\begin{document}

\title{Majorana oscillations {modulated} by Fano interference {and dregree of non-locality} in a topological superconducting nanowire-quantum dot system}

\author{L. S. Ricco}
\affiliation{Departamento de F\'{i}sica e Qu\'{i}mica, Unesp - Univ
Estadual Paulista, 15385-000, Ilha Solteira, S\~ao Paulo, Brazil}
\author{V. L. Campo Jr.}
\affiliation{Departamento de F\'isica, Universidade Federal de S\~ao Carlos, Rodovia Washington Luiz, km 235, Caixa Postal 676, 13565-905, S\~ao Carlos, S\~ao Paulo, Brazil}
\author{I. A. Shelykh}
\affiliation{Science Institute, University of Iceland, Dunhagi-3, IS-107,
Reykjavik, Iceland}
\affiliation{ITMO University, St. Petersburg 197101, Russia}
\author{A. C. Seridonio}
\email[correspondent author: ]{acfseridonio@gmail.com}
\affiliation{Departamento de F\'{i}sica e Qu\'{i}mica, Unesp - Univ
Estadual Paulista, 15385-000, Ilha Solteira, S\~ao Paulo, Brazil}
\affiliation{Instituto de Geoci\^encias e Ci\^encias Exatas - IGCE,
Universidade Estadual Paulista, Departamento de F\'{i}sica, 13506-970,
Rio Claro, S\~ao Paulo, Brazil}

\date{\today}

\begin{abstract}
We explore theoretically {the influence of Fano interference in} the so-called Majorana oscillations in a {T-shaped} hybrid setup formed by a quantum dot (QD) placed between conducting leads and side-coupled to a topological superconducting nanowire {(TSNW)} hosting zero-energy Majorana bound states (MBSs) at the ends. Differential conductance as a function of the external magnetic field reveals oscillatory behavior. Both the shape and amplitude of the oscillations depend on the bias-voltage, {degree of MBSs non-locality} and Fano parameter of the system determining the regime of interference. When the latter is such that direct lead-lead path dominates over lead-QD-lead path and the bias is tuned in resonance with QD zero-energy, pronounced {fractional Fano-like resonances are observed around zero-bias for highly non-local geometries. Further, the conductance profiles as a function of both bias-voltage and QD energy level display \textit{``bowtie''} and \textit{``diamond''} shapes, in qualitative agreement with both previous theoretical and experimental works. These findings ensure that our proposal can be used to estimate the degree of MBS non-locality, thus allowing to investigate their topological properties.}

\end{abstract}
\maketitle

\section{Introduction}\label{Introduction}

\begin{figure}[t]
	\centerline{\includegraphics[width=3.5in,keepaspectratio]{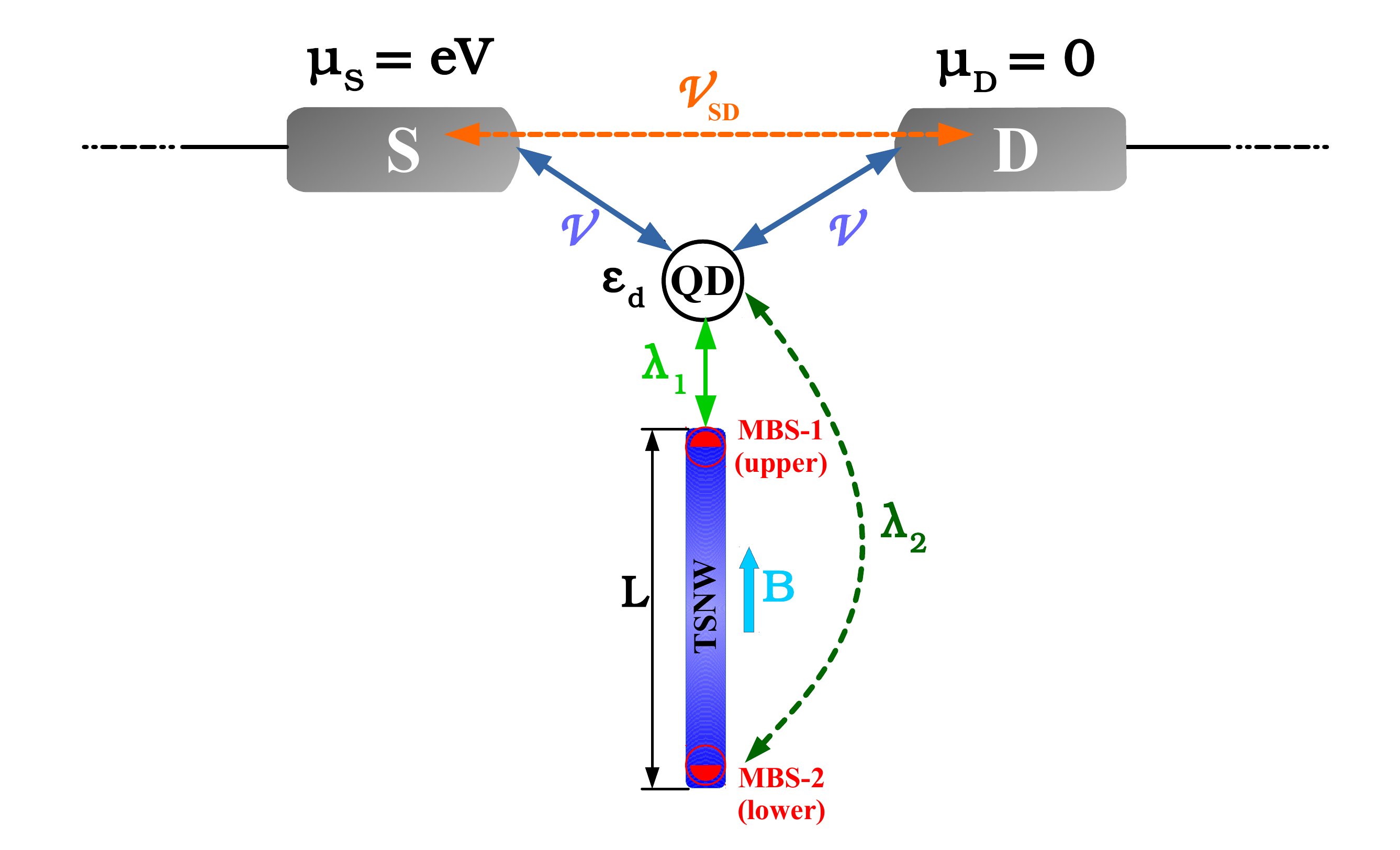}}
	\caption{Sketch of the {T-shaped} geometry considered in the present paper: a single-level QD {with energy $\varepsilon_{d}$} is hybridized symmetrically ($\mathcal{V}$) with source-drain (S/D) conducting leads and side-coupled ($\lambda_{1}$) to a TSNW of length $L$, hosting zero-energy MBSs at the edges {(half-filled red circles). The coupling $\lambda_{2}$ between the QD and MBS-2 is also taken into account due to the finite length of topological nanowire.} Leads are also coupled with each other directly ($\mathcal{V}_{SD}$). External magnetic field $B$ (light blue arrow) is applied parallel to the direction of the wire. The value of Zeeman splitting induced by magnetic field is considered to be large enough to achieve the full spin polarized regime in the setup. {The QD-leads system operates as a tunneling spectrometer, allowing to investigate the properties of the MBSs by differential conductance measurements} as a function of the external magnetic field and bias-voltage between leads. {The presence of lead-lead tunneling path allows to explore how Fano interference process affects the MBSs signatures.}
		\label{Setup}}
\end{figure}

Ideas borrowed from high energy physics became ubiquitous in the domain of condensed matter. The concepts of quasi-relativistic particles in graphene and other Dirac materials, acoustic analogs of black holes in Bose-Einstein condensates, AdS-CFT duality in the theory of the quantum phase transitions are now among standard tools used by condensed matter specialists. Some of these concepts still remain playground for theoreticians; the others on the contrary appeared to be of high experimental relevance and even paved way to novel applications in the domains of nanoelectronics and quantum computing. Among these latter are Majorana quasiparticles ~\cite{alicea2012new,RevMajorana} which are currently considered as highly perspective candidates for practical realization of fault-tolerant quantum computation process~\cite{RevNonabelian}.

In the domain of condensed matter, Majorana quasiparticles appear in hybrid systems composed by a quasi-one-dimensional semiconducting nanowire with strong spin-orbit coupling placed nearby \textit{s}-wave superconductors~\cite{Nanowire1, Nanowire2, Nanowire3, Kitaev2001unpaired}. In this configuration, when magnetic (Zeeman) field is applied parallel to the wire, the latter enters into \textit{p}-wave topological superconducting (SC) phase and a pair of gapless (zero-energy) Majorana bound states (MBSs) is formed at the nanowire edges~\cite{Kitaev2001unpaired}. {To analyze the MBSs transport properties in hybrid systems with topological superconducting nanowire (TSNW), in several theoretical works was proposed the use of quantum dots(QDs) as tunneling spectrometers to reveal MBSs signatures and topological transitions ~\cite{baranger,JAP,leakage,DJClarcke,RAguado,Recher} . Experiments on hybrid TSNW with a QD also were performed~\cite{wire2016,Aguadoexp}.} According to Liu and Baranger prediction~\cite{baranger}, the experimental signature of the onset of an isolated MBS is zero-bias peak (ZBP) with $e^{2}/2h$ amplitude in the conductance profile of a system consisting of an individual QD side-coupled to a TSNW.

{Despite experimental observations of the quantized ZBP in sophisticated devices with hybrid TSNWs ~\cite{mourik2012,albrecht2016exponential,wire2016,epitaxial,Aguadoexp,Ballistic,Quantized},} questions regarding its amplitude and emergence inside a truly topological phase still remain~\cite{Jelena,PhysRevLett.109.267002,PhysRevB.93.245404,PhysRevB.96.075161,DJClarcke,RAguado}. {Indeed, only the observation of ZBP itself is not sufficiently for asserting that the system is into the topological regime, hosting robust MBSs. In this context, non-local Majorana features~\cite{DJClarcke,RAguado}, as well as ZBP splitting, followed by the appearance of an oscillatory pattern in the differential conductance as function of an applied magnetic field~\cite{DasSarma,Jelena,PhysRevB.96.125420} have been viewed as smoking guns of the MBSs manifestation in the topologically non-trivial regime.}

In the current work, we investigate {the role of Fano interference processes in the so-called Majorana oscillations} in a {T-shaped} nanodevice, consisting of a single-level QD placed between the conducting leads and side-coupled to a TSNW hosting MBSs at the ends (Fig.~\ref{Setup}). {We also explore features of the system when the Majorana non-locality is taken into account, by considering the coupling $ \lambda_{2} $ between the QD and lower MBS (see Fig.~\ref{Setup}). The degree of Majorana non-locality $\eta$ was previously defined by Prada \textit{et al.}~\cite{RAguado} as the ratio between the lower and upper QD-MBSs coupling strengths, \textit{i.e.}, $\eta^{2} = |\lambda_{2}|/|\lambda_{1}|$. When $\eta \rightarrow 0~(|\lambda_{2}| \ll |\lambda_{1}|)$, the MBSs are highly non-local, thus presenting the holy grail for the quantum computation: the topological protection feature. Ref.~[\onlinecite{RAguado}] also proposed a protocol to estimate experimentally such a degree of non-locality, which was recently performed by Deng \textit{et al.}~\cite{Aguadoexp} in a TSNW, with a QD working as spectrometer. The ratio between QD-MBSs couplings also can define the \textit{``topological quality factor''}, being stated by Clarke~\cite{DJClarcke} as $\mathcal{Q} = 1 - \eta^{2}$. In this context, the higher topological quality occurs when $\mathcal{Q} \approx 1$.}

{It is worth mentioning that T-shaped setups with QDs are suitable geometries to investigate the well-known Fano effect~\cite{Fano,RevFano,FanoTshaped,Fanoparameter}, once they have the key ingredients for its emergence: a localized state coupled to the continuum and distinct tunneling channels. Fano interference phenomenon can be used to explore Majorana properties, as theoretically proposed in earlier works~\cite{JAP1,JAP,Recher,Baranski}.}

In the current proposal, the quantities which define the tunneling conductance spectroscopy are Zeeman field, bias-voltage between the leads, energy level of QD, couplings between the MBSs and QD and Fano parameter, describing the relative importance of the direct lead-lead and lead-QD-lead tunneling paths. The conductance as a function of the magnetic field reveals pronounced oscillatory pattern, {which are both dependent on Fano regime of interference and MBSs non-local features. In a nutshell, when the direct lead-lead tunneling prevails, the Majorana oscillations are suppressed at zero-bias and reveal unexpected fractional Fano-like resonances as a function of bias-voltage between the leads. The degree of MBSs non-locality also influences the behavior of such oscillations, which are attenuated as the local feature is increased (lower topological quality factor). We also report the ability to identify experimentally such a degree of non-locality in conductance measurements by changing the energy level of QD. Our results are in agreement with Ref.~[\onlinecite{RAguado}], despite differences between their system and ours, which will be discussed in due course.}

{This work is organized as follows: in Sec.~\ref{model} we present the theoretical model describing the system of Fig.~\ref{Setup}. We also show the expression for zero-bias conductance and corresponding transmittance through the QD, which was obtained via equation of motion (EOM) technique. In Sec.~\ref{Results} we show and discuss our findings, which are summarized in Sec.~\ref{conclusions}}.

\section{The Model}\label{model}

The setup we consider is depicted in Fig.~\ref{Setup} and can be described by the following spinless model Hamiltonian~\cite{baranger}:

\begin{eqnarray}
\mathcal{H} & = &\sum_{\alpha,k}\xi_{\alpha,k}c_{\alpha,k}^{\dagger}c_{\alpha,k}+\varepsilon_{d}d^{\dagger}d+ \mathcal{V}\sum_{\alpha,k}(c_{\alpha,k}^{\dagger}d+\text{H.c}) \nonumber \\
& + &  \mathcal{V}_{SD} \sum_{k,l}(c_{S,k}^{\dagger}c_{D,l}+\text{H.c}) + \mathcal{H}_{\text{M}}, \label{eq:H}
\end{eqnarray}
where the operator $c_{\alpha,k}^{\dagger}(c_{\alpha,k})$ creates
an electron (hole) in the metallic lead $\alpha=S/D$ (Source/Drain)
with wave-number $k$ and energy $\xi_{\alpha,k}=\varepsilon_{k}-\mu_{\alpha}$,
where $\mu_{\alpha}$ is chemical potential and $\mu_{S}-\mu_{D}=eV$
is the bias-voltage between the leads. The operator $d^{\dagger}(d)$ creates
an electron (hole) in the energy level $\varepsilon_{d}$ of
the QD, which is symmetrically coupled to the leads with coupling constant $\mathcal{V}$. The lead-lead coupling constant is $\mathcal{V}_{SD}$. {No charging effect was taken into account in the QD energy level, since the MBS signatures remain in presence of Coulomb repulsion and possible Kondo physics, as Ruiz-Tijerina \textit{et al.}\cite{Tijerina} have shown.}

{Considering the even and odd conduction operators $c_{e,k} = c_{S,k}\cos \theta + c_{D,k} \sin \theta$ and $c_{o,k} = c_{S,k}\sin \theta - c_{D,k} \cos \theta$, with $\tan \theta = 1$, Eq.~(\ref{eq:H}) can be rewritten as}
{
\begin{eqnarray}
\mathcal{H} & = &\sum_{k}\varepsilon_{k} c^{\dagger}_{e,k}c_{e,k} + \varepsilon_{d}d^{\dagger}d + \sqrt{2}\mathcal{V} \sum_{k} (c_{e,k}^{\dagger}d+\text{H.c})\nonumber \\
& + & \mathcal{V}_{SD} \sum_{k,q} c_{e,k}^{\dagger}c_{e,q} + \mathcal{H}_{\text{M}} + \mathcal{H}_{o}, \label{eq:even-odd}
\end{eqnarray}}{wherein $\mathcal{H}_{o} = \sum_{k}\varepsilon_{k}c^{\dagger}_{o,k}c_{o,k} - \mathcal{V}_{SD}\sum_{k,q}c_{o,k}^{\dagger}c_{o,q}$ describes the odd conduction states, which are decoupled from the QD~\cite{JAP}.}

The term~\cite{SciReports}
\begin{equation}
\mathcal{H}_{\text{M}}=\imath\varepsilon_{M}\gamma_{1}\gamma_{2}+\lambda_{1}(d-d^{\dagger})\gamma_{1} + {\lambda_{2}(d + d^{\dagger})\gamma_{2}}\label{eq:2}
\end{equation}
is the effective model Hamiltonian for a TSNW hosting zero-energy MBSs $\gamma_{i}$ at the ends~\cite{Kitaev2001unpaired}. The Majorana operators have following algebra ~\cite{alicea2012new}: $\left[\gamma_{i}, \gamma_{j} \right]_{+} = \delta_{ij}$, $\gamma_{i(j)}^{\dagger}=\gamma_{i(j)}$.
The parameter $\varepsilon_{\text{M}}\equiv\varepsilon_{\text{M}}(l,B)=\frac{E_{0}}{\sqrt{b}}e^{-l/2b}\cos(l\sqrt{b})$ describes the overlapping of unpaired gapless MBSs at the opposite sides of the wire~\cite{PhysRevB.96.125420}, where $b=B/E_{0}$, $l=L\sqrt{2mE_{0}}/\hbar$ with $B$ being longitudinal Zeeman field (light-blue arrow in the Fig.~\ref{Setup}), $L$ the length of the wire, $E_{0}=(2m\alpha^{2}\Delta_{\text{SC}}^{2}/\hbar^{2})^{1/3}$~\cite{PhysRevB.96.125420}, $\alpha$ is spin-orbit constant and $\Delta_{\text{SC}}$ is the induced SC gap in the wire. The presence of the term $\cos(l\sqrt{b})$ in $\varepsilon_{\text{M}}$ is responsible for the oscillatory pattern in conductance as function of the magnetic field. {The couplings between the upper/lower MBSs and the QD are given by $\lambda_{1}$ and $\lambda_{2}$, respectively~\cite{SciReports}. As known, the Hamiltonian of Eq.~(\ref{eq:2}) can be rewritten with usual fermion operators~\cite{alicea2012new} $f$, since $\gamma_{1}=\frac{1}{\sqrt{2}}\left(f+f^{\dagger}\right)$ and
$\gamma_{2}=\frac{\imath}{\sqrt{2}}\left(f^{\dagger}-f\right)$. To see how $\mathcal{H}_{\text{M}}$ stays in the fermionic basis, please see Ref.~[\onlinecite{SciReports}].}

The differential conductance of the system is given by the following expression~\cite{Jauho}:

\begin{equation}
\mathcal{G}(eV) =  \frac{e^{2}}{h}\int \left(- \frac{\partial f_{F}(\omega,eV)}{\partial \omega} \right)\mathcal{T}(\omega) d \omega=\frac{e^{2}}{h} \mathcal{T}(eV)\label{eq:zbconductance},
\end{equation}
where $e^{2}/h$ is quantum of conductance and $f_{F}$ is Fermi-Dirac distribution function. The last equality holds for $T=0$. $\mathcal{T}(\omega) $ is the transmittance across the system which can be obtained using equation of motion (EOM) method \cite{Jauho,DESSOTTI2016297}, yielding:

\begin{eqnarray}
\mathcal{T}(\omega) & = & \mathcal{T}_{b} + \sqrt{\mathcal{T}_{b} \mathcal{R}_{b}} \tilde{\Gamma} \text{Re}\left[ G^{r}_{dd}(\omega) \right]\nonumber \\
& - & (1 - 2\mathcal{T}_{b})\frac{\tilde{\Gamma}}{2}\text{Im}\left[ G^{r}_{dd}(\omega) \right], \label{eq:Transmittance}
\end{eqnarray}
where $\tilde{\Gamma} = \Gamma / (1+x)$ is dot-lead effective coupling,
$\Gamma = 2\pi \mathcal{V}^{2}\sum_{k}\rho_0$ is Anderson broadening~\cite{Anderson},  $x=(\pi\mathcal{V}_{SD}\rho_0)^2$, $\rho_0=\sum_{k}\delta(\omega - \varepsilon_{k})$ is the density of states (DoS) of the leads, $\mathcal{T}_{b} = 4x/(x+1)^{2}$
and $\mathcal{R}_{b} = 1- \mathcal{T}_{b}$ are the background transmittance and reflectance, respectively~\cite{JAP,PhysRevB.94.125426}. We also define the Fano parameter~\cite{Fanoparameter} $q_{b} = \sqrt{\frac{\mathcal{R}_{b}}{\mathcal{T}_{b}}} = \frac{(1-x)}{2\sqrt{x}}$. {For asymmetric couplings between the QD and leads~\cite{Vivaldo}, $\mathcal{H}_{o}$ of Eq.~(\ref{eq:even-odd}) remains decoupled from the QD, with $\tan \theta = \mathcal{V}_{S}/\mathcal{V}_{D}$. The only differences are an effective Anderson broadening $\Gamma'  = 2 \Gamma_{S} \Gamma_{D}/(\Gamma_{S} + \Gamma_{D})$ and an effective QD-even conduction band coupling  $\mathcal{V}' = \sqrt{\mathcal{V}_{S}^{2} + \mathcal{V}_{D}^{2}}$ instead of $\sqrt{2}\mathcal{V}$.}

To calculate the spectral retarded Green's function of the QD $G^{r}_{dd} (\omega)$ in the Eq.~(\ref{eq:Transmittance}), we use again EOM technique, which allows us to get the following expression:

\begin{equation}
 G^{r}_{dd}(\omega) = \frac{1}{\omega^{+} - \varepsilon_{d} - \Sigma - \Sigma_{\text{MBSs}}(\omega)}, \label{Gdd}
\end{equation}
where $\Sigma = -(\sqrt{x} + \imath)\Gamma / (1+x)$, {$\Sigma_{\text{MBSs}}(\omega) = K_{+}(\omega) + (|\lambda_{1}|^{2} - |\lambda_{2}|^{2})\tilde{K}(\omega)K(\omega)$} is the part of self-energy provided by the presence of MBSs \cite{baranger,JAP}, {$\tilde{K}(\omega) = K(\omega)/ (\omega^{+} + \varepsilon_{d} + \Sigma^{*} - K_{-}(\omega))$}, $K(\omega) = \omega^{+} / [(\omega^{+})^{2} - \varepsilon_{\text{M}}^{2}]$ and
\begin{equation}
{
K_{\pm}(\omega) = \frac{\omega^{+}(|\lambda_{1}|^{2} + |\lambda_{2}|^{2}) \mp 2\varepsilon_{\text{M}}|\lambda_{1}||\lambda_{2}|}{[(\omega^{+})^{2} - \varepsilon_{\text{M}}^{2}]},\label{eq:Kpm}}
\end{equation}
{with $\omega^{+} = \omega + \imath 0^{+}$.} Imaginary part of the Green's function given by Eq.~(\ref{Gdd}) defines the DoS of the QD,
\begin{equation}
\rho_{dot}(\omega)= -\frac{1}{ \pi} \text{Im} [ G^{r}_{dd}(\omega)].
\end{equation}

\section{Results and Discussion}\label{Results}

 We investigate the effects of applied longitudinal Zeeman field on differential conductance of the system restricting ourselves to the temperature $T=0$. Our goal is to analyze the changes in the conductance oscillation patterns introduced by the bias-voltage between leads {for distinct Fano regimes of interference and couplings between the QD and MBSs. The tuning of QD-lower MBS coupling strength $|\lambda_{2}|$ allows to study the degree of MBS non-locality $\eta$, as discussed in Sec.~\ref{Introduction}}. Concerning the Fano interference process, one should discriminate between the cases when tunneling between the leads goes preferably via QD as intermediate ($x=0$, $q_{b} \rightarrow \infty$, $\mathcal{T}_{b}= 0$) and the opposite case when direct lead-lead tunneling prevails ($x = 1$, $q_{b} = 0$ $\mathcal{T}_{b} = 1$). {Intermediary situations are also considered ($0<x<1$)}. The parameters of the system are taken as: $E_{0}\approx \SI{0.23}{\milli\electronvolt}$ the wire length $L = \SI{1}{\micro\meter}$, and Anderson broadening $\Gamma \approx 0.17E_{0}$. This is in agreement with both experiment~\cite{mourik2012,albrecht2016exponential,wire2016} and existing theoretical estimations~\cite{DasSarma,leakage}.

 {Before discussing in detail our findings, we define the Zeeman critical value $B_{c}$, corresponding to $b=(\pi/2l)^{2}$, as the value in which MBSs begin to overlap with each other. It is important to mention that the Hamiltonian which describes the system [Eq.~(\ref{eq:H})] is an effective model that previously takes into account a Zeeman field to break the spin degeneracy, thus ensuring the spinless feature considered here and appearance of MBSs. Besides this field, intrinsic to the model, there is an applied longitudinal Zeeman field $B$ in the TSNW, which overlaps the MBSs for $B>B_{c}$ and is responsible for oscillatory pattern in the conductance, as we shall see.}


\subsection{{Majorana Oscillations and Fano interference}}\label{oscillations}

\begin{figure}[t]
\centerline{\includegraphics[width=3.5in,keepaspectratio]{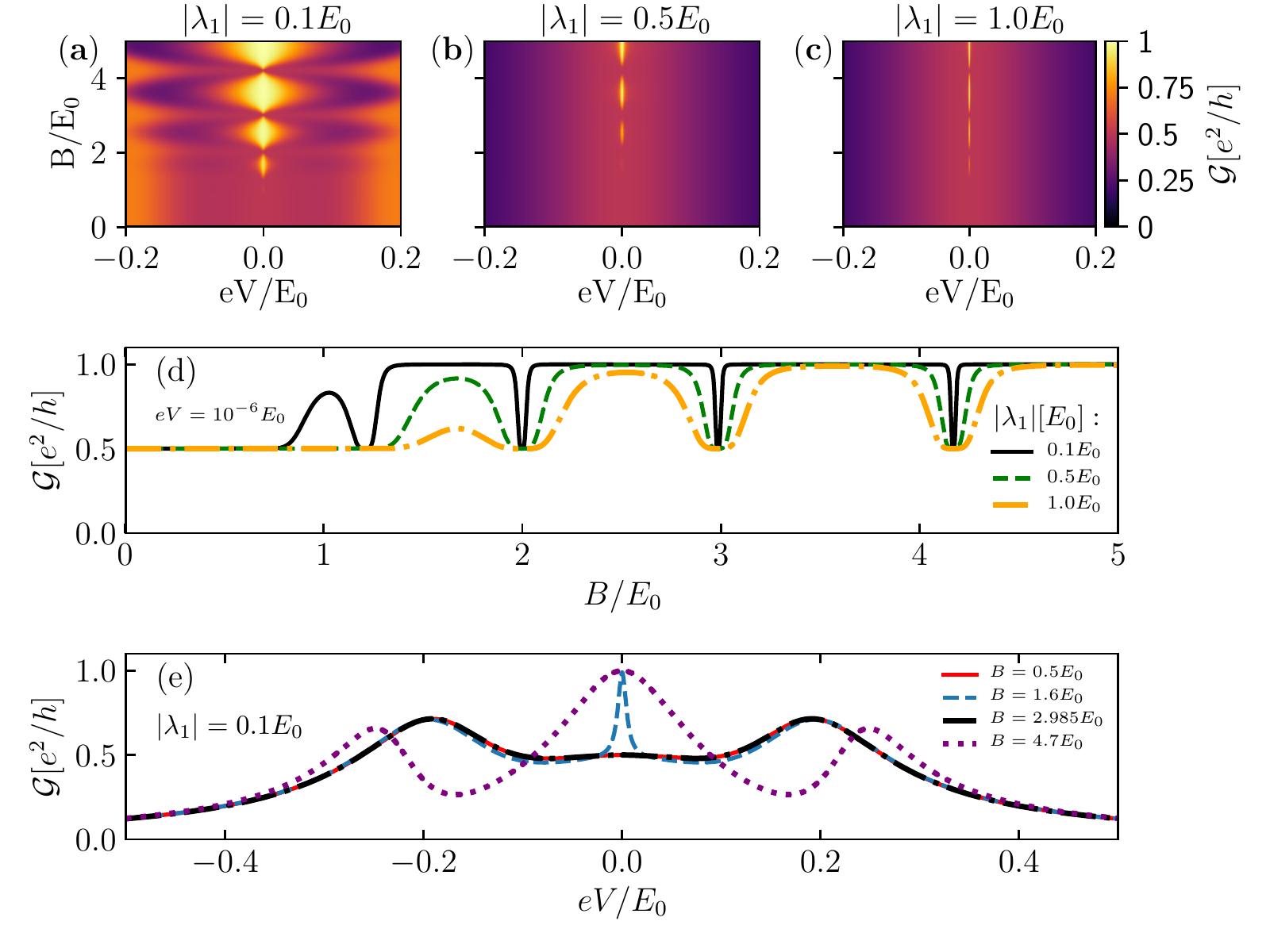}}
\caption{(a)-(c): Plots of differential conductance [Eq.~(\ref{eq:zbconductance})] as a function of the Zeeman field $B$ and bias-voltage $eV$ for the case when lead-QD-lead tunneling path is dominant ($x=0, q_{b}\rightarrow \infty$) and {$|\lambda_{2}|=0$}. The QD energy level $\varepsilon_d=0$, the values of TSNW-QD coupling are $|\lambda_{1}|=0.1E_0,0.5E_0$ and $E_0=1.0E_0$. (d) Differential conductance at $eV=10^{-6}E_0$ as function of the Zeeman field, for distinct values of TSNW-QD coupling $|\lambda_1|$. (e) Differential conductance as a function of $eV$ for several values of the Zeeman field.
\label{Result1}}
\end{figure}

{In this section, we study the role of Fano interference effect in the Majorana oscillations emerging in differential conductance, for $\varepsilon_{d}=0$. Fig.~{\ref{Result1}} shows the differential conductance as a function of both $eV$ and Zeeman field, considering different TSNW-QD couplings$(|\lambda_{1}|)$ for the case when lead-QD-lead tunneling path is dominant ($x=0, q_{b}\rightarrow \infty$) and $|\lambda_{2}|=0$. In such a case the MBSs can overlap via $\varepsilon_{\text{M}}$, but the wire is long enough to ensure that there is no connection between the MBS-2 and QD (See Fig.~\ref{Setup})}. If the value of magnetic field is below critical, $B<B_{c}$, we observe typical plateau in differential conductance with $\mathcal{G}=e^{2}/2h$, which indicates that MBSs remain isolated from each other. When magnetic field exceeds the critical value, $B>B_{c}$, an oscillatory pattern in differential conductance as function of the magnetic field arises. The value of the conductance oscillates between the minimal value of $e^{2}/2h$ and maximal value which in certain cases can reach $e^{2}/h$. This latter points to a regular fermion signature arising due to the finite overlap between the MBSs~\cite{baranger}.

These effects become visible at the panel (d), where the differential conductance is plotted as function of the magnetic field for $eV=10^{-6}E_0$: the oscillations between isolated MBSs ($\varepsilon_{\text{M}} \rightarrow 0$, $\mathcal{G}\rightarrow e^2/2h$) and nonlocal fermion state formed by overlapping MBSs ($\mathcal{G}\rightarrow e^2/h$) are clearly visible. The increase of TSNW-QD coupling $\lambda_1$ broadens the dips in the conductance and decreases the amplitudes of the oscillations. Note however, that for big values of magnetic field the maxima of the conductance still reach the values of the conductance quantum. Effects of the overlap between MBSs assisted by Zeeman field are clearly seen at panel (e), where differential conductance is plotted versus $eV$. Indeed, for certain values of $B$ (corresponding e.g. to dashed-blue and dotted purple lines), conductance reaches maximum value at $eV=0$, which is a signature of a regular fermion state, whereas for other values of Zeeman field (corresponding e.g. to red filled and dash-dotted black lines), $\mathcal{G}$ has minima at $eV=0$, which corresponds to the case of isolated MBSs.

\begin{figure}[t]
\centerline{\includegraphics[width=3.5in,keepaspectratio]{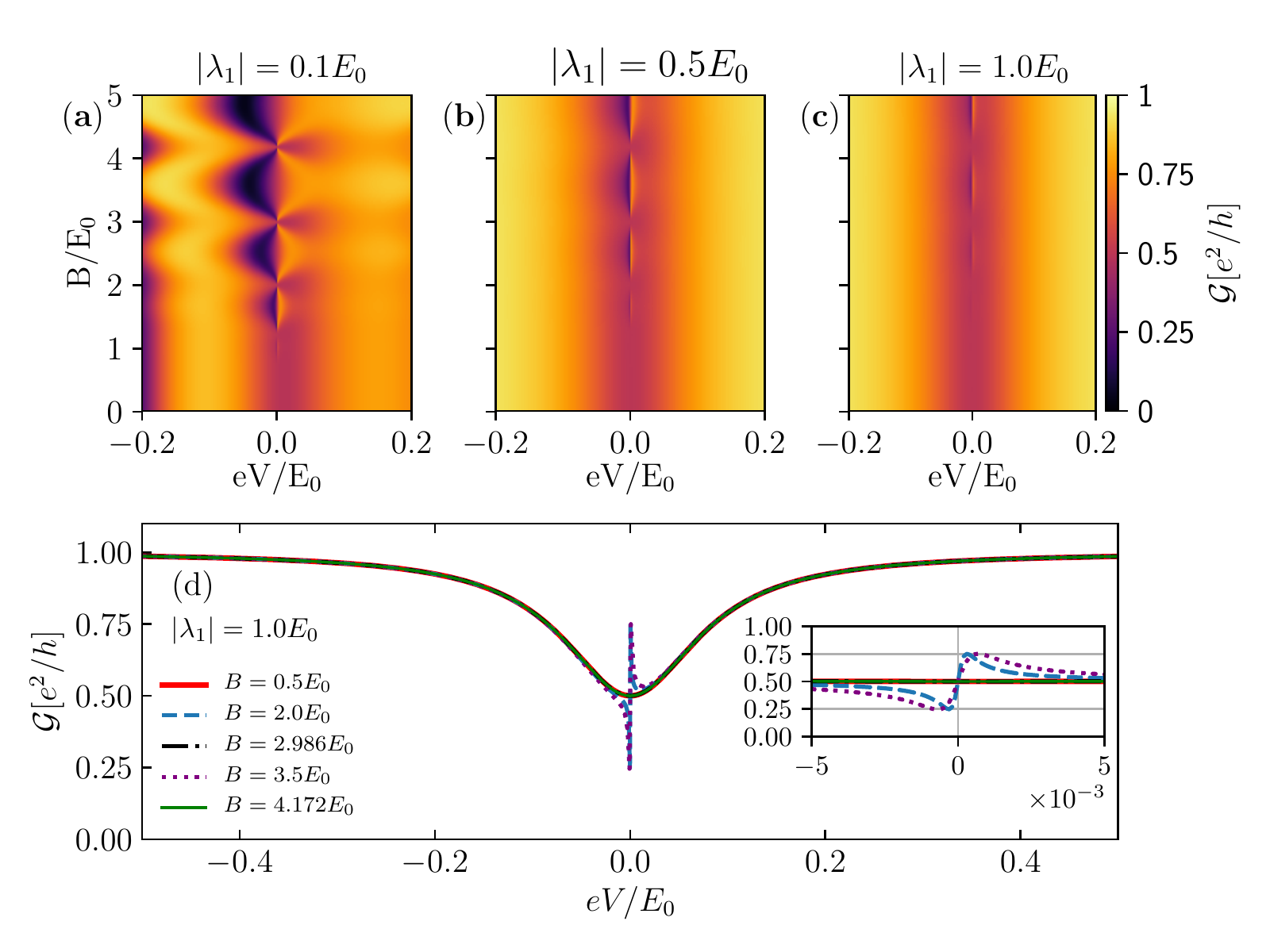}}
\caption{(a)-(b): Plots of differential conductance [Eq.~(\ref{eq:zbconductance})] as function of the Zeeman field $B$ and $eV$, for the case when direct lead-lead tunneling path is dominant ($x=1, q_{b} = 0$) and {$|\lambda_{2}| = 0$}. The QD energy level $\varepsilon_d=0$, the values of TSNW-QD coupling are $|\lambda_{1}|=0.1E_0,0.5E_0,$ and $E_0=1.0E_0$. (d) Differential conductance as function of $eV$ for several values of Zeeman field. Conductance reveals sharp resonant asymmetric profile.
\label{Result3}}
\end{figure}

The change of the Fano interference regime to the case where $q_{b} = 0$, corresponding to the dominance of the direct lead-lead tunneling, brings dramatic changes in the differential conductance pattern, as can be seen at Fig.~\ref{Result3}. Two qualitatively new phenomena are observed here as compare to the case $q_{b}\rightarrow \infty$. First, at $eV=0$, $\mathcal{G}=e^{2}/2h$ and is independent on the values of $|\lambda_{1}|$ and applied field $B$. Moreover, differential conductance as function of bias-voltage reveals {fractional Fano-like resonances around} $eV=0$ with intriguing minimal and maximal values equal to $e^{2}/4h$ and $3e^{2}/4h$ ~\cite{PhysRevB.94.125426,JAP1}. {Similar fractional Fano interference process was already reported by Bara\'{n}ski \textit{et al.}~\cite{Baranski} in a T-shaped geometry with a QD between metallic and superconducting leads, side-coupled to a MBS. In such a system, the fractional interferometric behavior is related to the presence of MBS in the system, which scatters the electron waves, changing their phase~\cite{Baranski}. We highlight that the fractional oscillatory pattern reported here only can be verified for low temperatures ($T \leq mK$). Otherwise, the thermal effects can smear out such Fano-like resonances, making the effect unobservable.}

\begin{figure}[t]
\centerline{\includegraphics[width=3.5in,keepaspectratio]{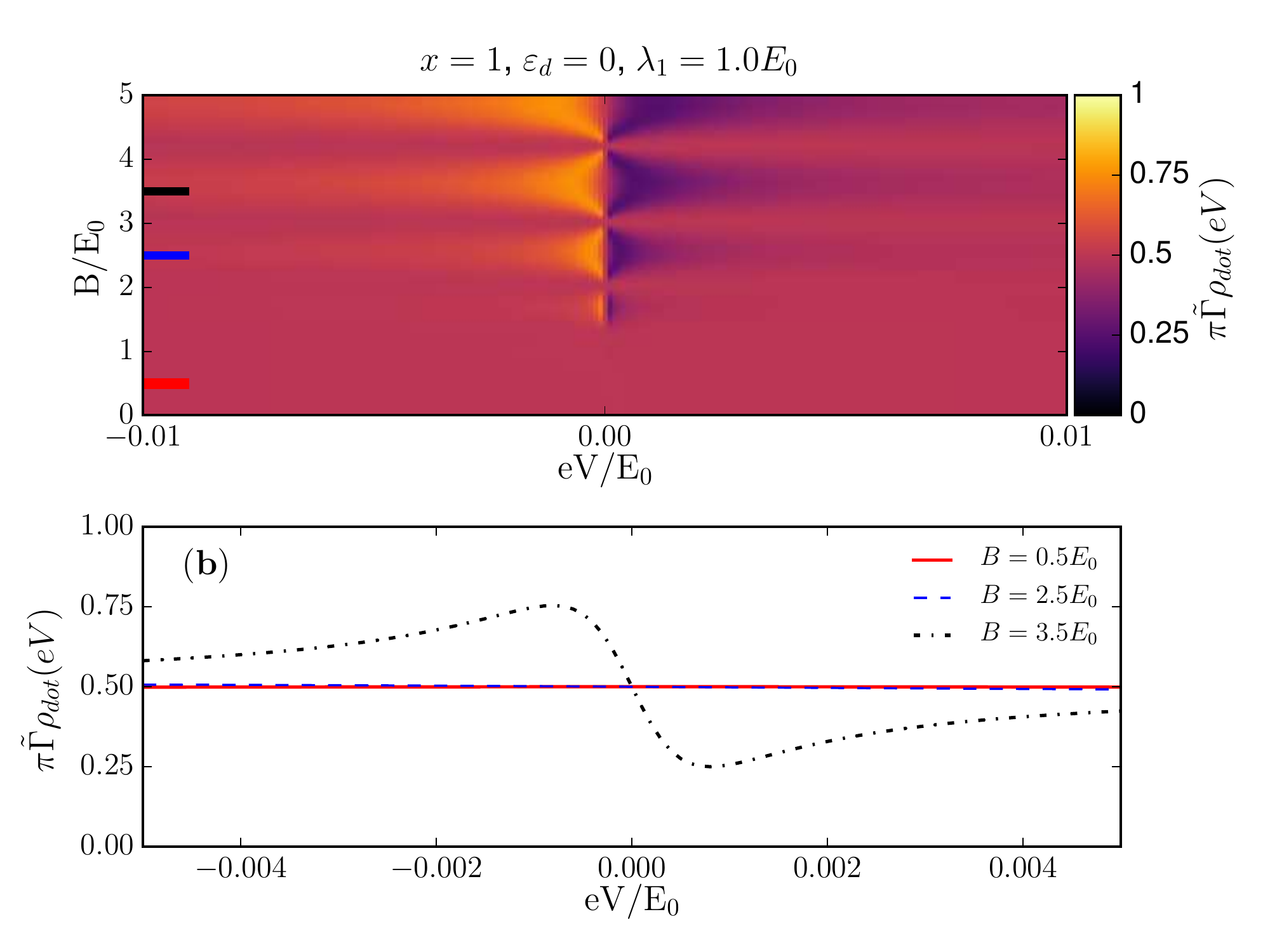}}
\caption{(a) Dimensionless DoS of the QD for the case of the dominant direct lead-lead tunneling ($x=1$, $q_{b}=0$) as a function of the Zeeman field and $eV$, with $|\lambda_{1}|=1.0E_{0}$ and $|\lambda_{2}|=0$. (b) Dimensionless DoS of the QD as function of $eV$ for three different values of the magnetic field, corresponding to the colored horizontal bars at the panel (a).
\label{Result4}}
\end{figure}

{We also examine the corresponding dimensionless QD DoS for the case $q_{b} = 0$.} The results are shown in the Fig.~\ref{Result4} for $|\lambda_{1}|=1.0E_{0}$. As it can be clearly seen, DoS reveal the resonant asymmetric pattern, which is inverted with respect to the pattern observed in the differential conductance: the dips in the DoS correspond to the peaks in $\mathcal{G}$ and vice versa. This inversion is a straight aftermath of the system electrical charge conservation: in the lead-lead Fano regime the better is localization of the electron on the dot the poorer is the conductance. In order to catch both charge conservation and {fractional Fano-like lineshapes}, we present horizontal line cuts of the color plot of the Fig.~\ref{Result4}(a) along red, blue and black horizontal bars, as shown in the Fig.~\ref{Result4}(b). As can be seen, for both values of $B>B_{c}$ considered (dashed blue and dash-dotted black lines), the amplitudes of {the fractional profile} remain the same.

\begin{figure}[t]
\centerline{\includegraphics[width=3.5in,keepaspectratio]{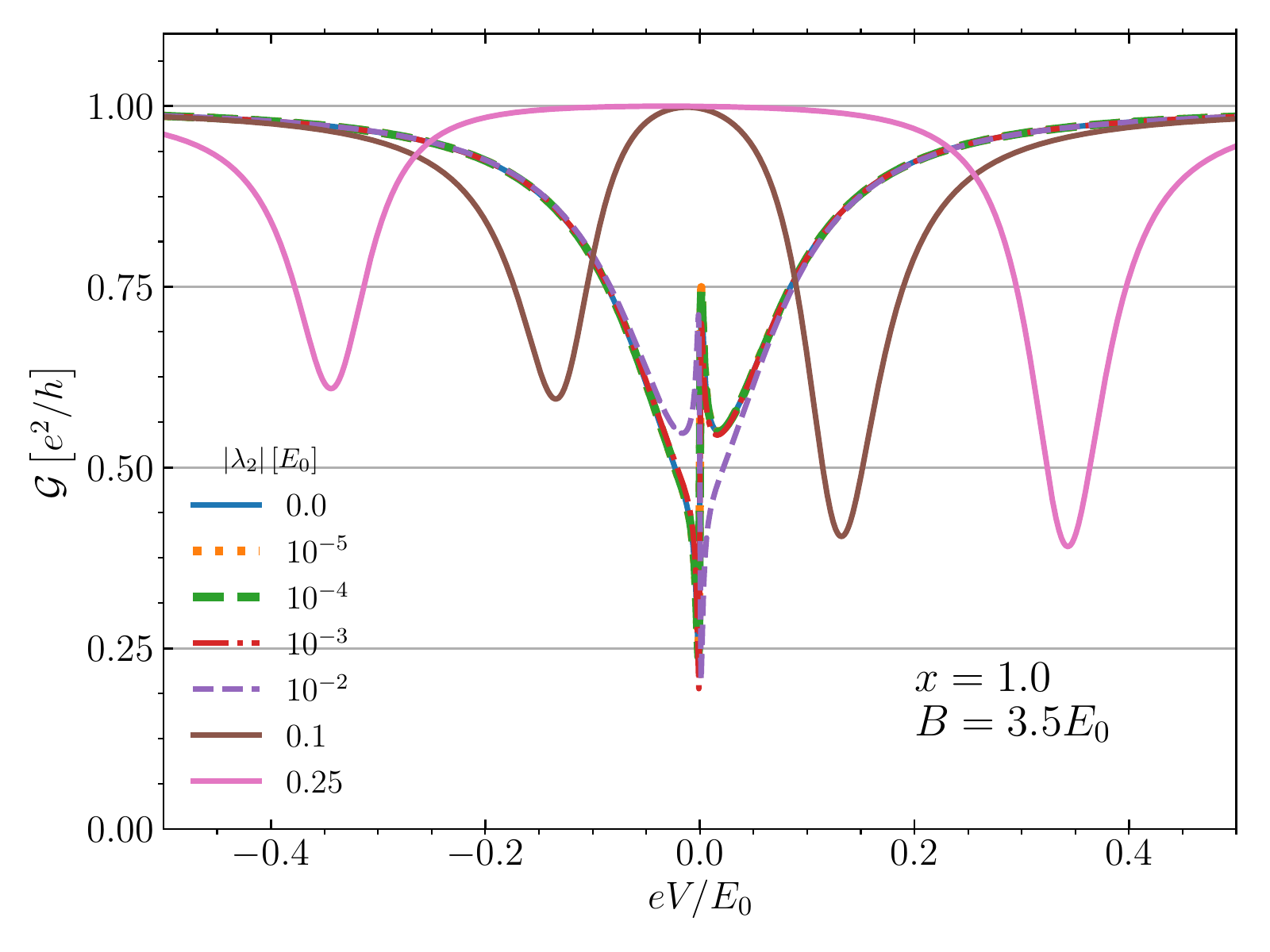}}
\caption{{Differential conductance [Eq.~(\ref{eq:zbconductance})] as function of $eV$  for the situation which the fractional Fano-like resonances are present($x=1$, $q_{b}=0$). Several values of the coupling between the QD and the lower MBS are considered($|\lambda_{2}|$).}
\label{Result5}}
\end{figure}

\begin{figure}[t]
\centerline{\includegraphics[width=3.5in,keepaspectratio]{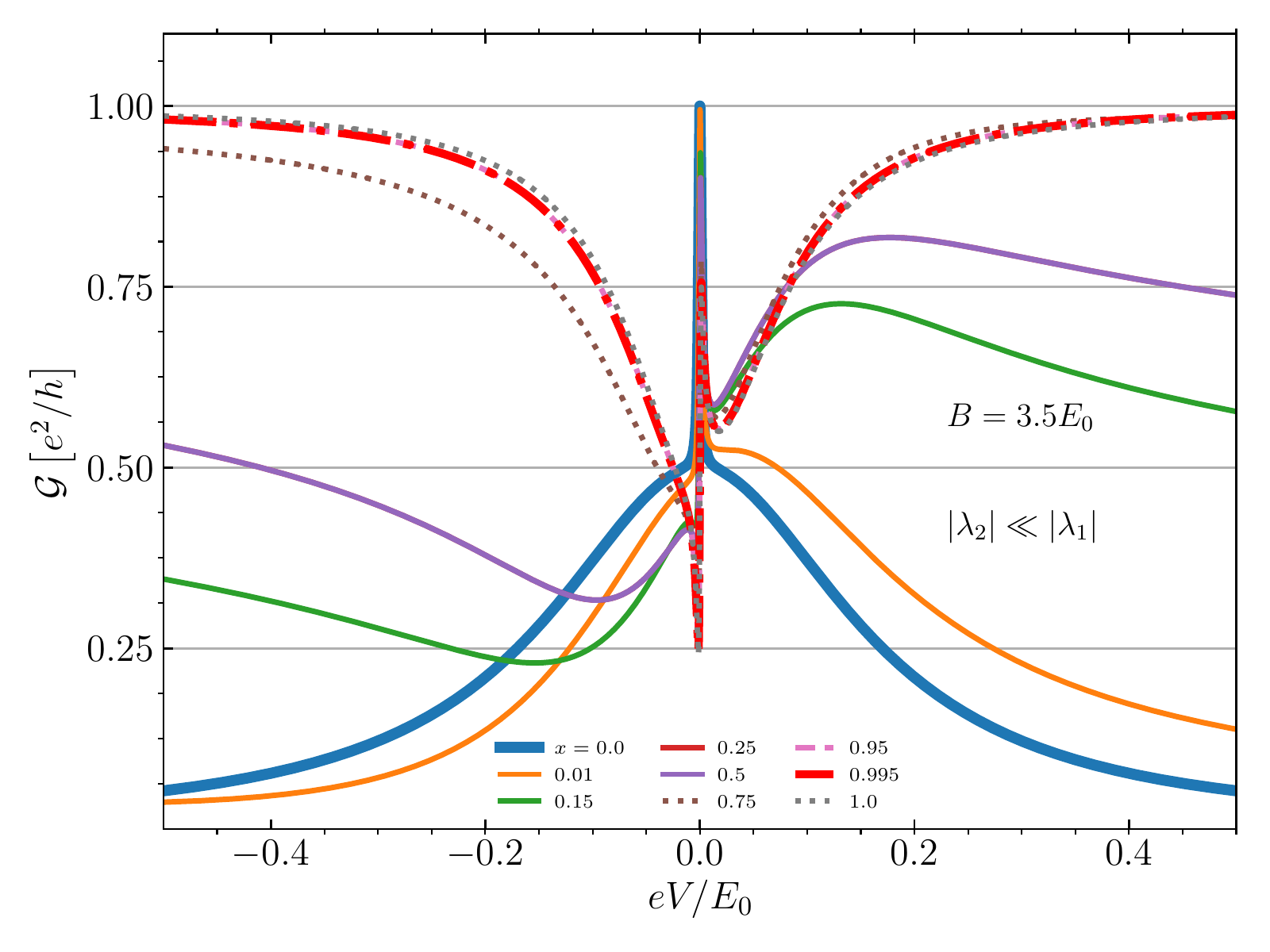}}
\caption{{Differential conductance [Eq.~(\ref{eq:zbconductance})] as function of $eV$ for the several Fano regimes of interference ($0\leq x \leq 1$) and highly non-local situation ($|\lambda_{2}| \ll |\lambda_{1}|$).}
\label{Result6}}
\end{figure}

\begin{figure}[t]
	\centerline{\includegraphics[width=3.5in,keepaspectratio]{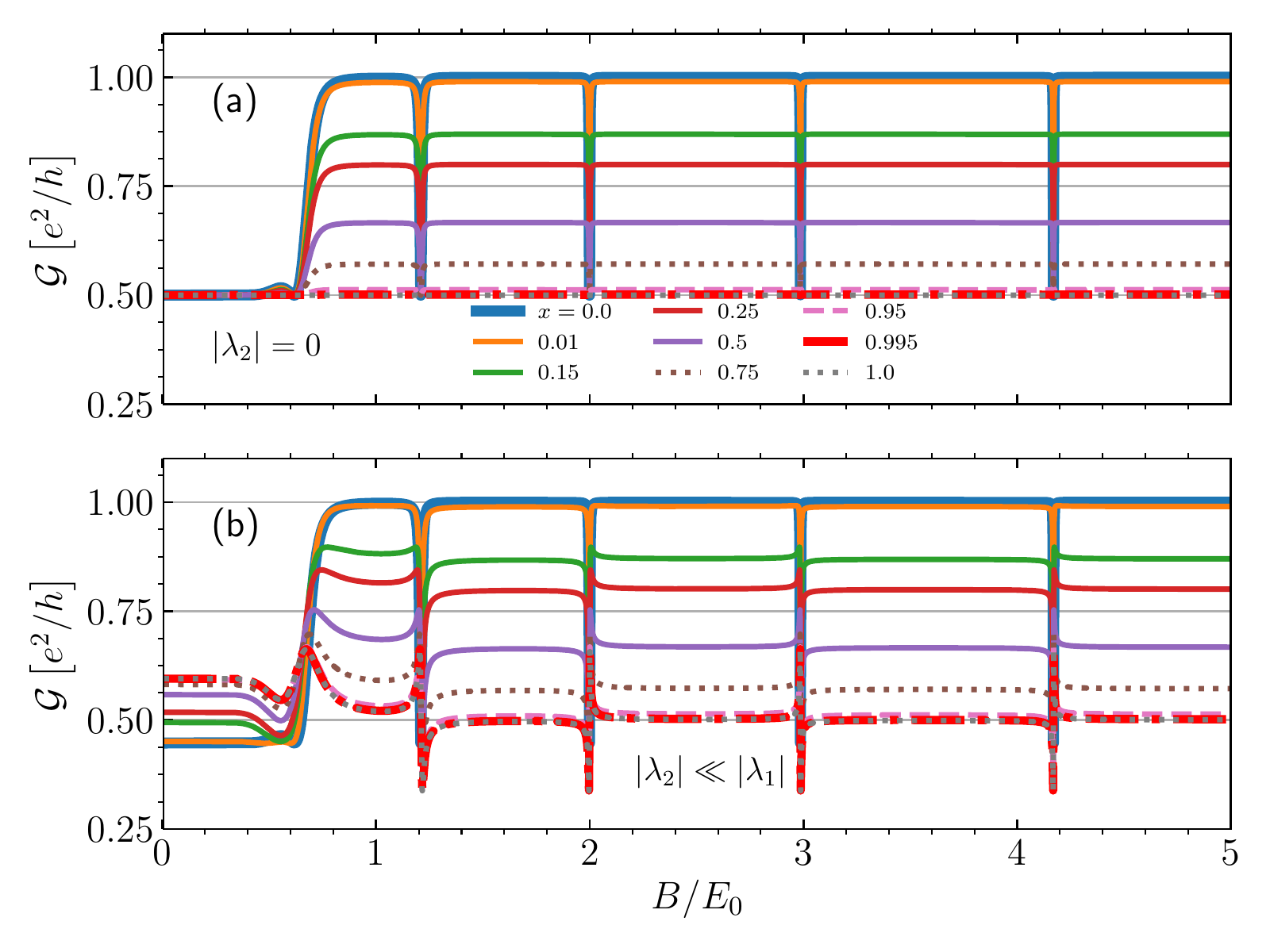}}
	\caption{{Differential conductance [Eq.~(\ref{eq:zbconductance})] as function of Zeeman field, for $eV=0$ and several Fano regimes of interference ($0\leq x \leq 1$). In panel (a) the coupling between the QD and the lower MBS is neglected $|\lambda_{2}=0|$, while in (b) is considered.}
		\label{Result7}}
\end{figure}

\begin{figure}[t]
	\centerline{\includegraphics[width=3.5in,keepaspectratio]{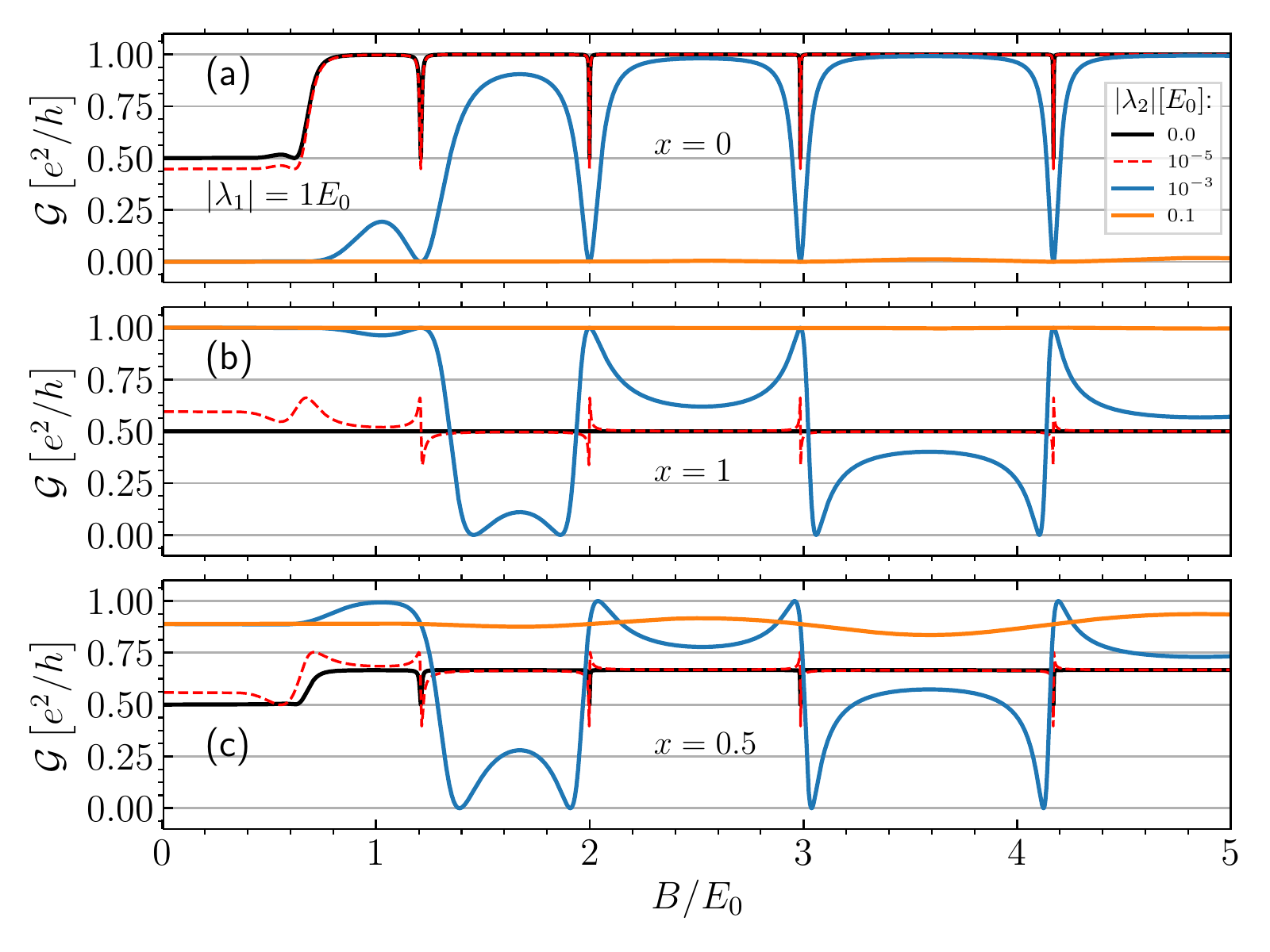}}
	\caption{{Differential conductance [Eq.~(\ref{eq:zbconductance})] as function of Zeeman field, for $eV=0$ and several values of $|\lambda_{2}|$. Panel (a) exhibits the Fano regime $x=0$, while (b) and (c) show the situation for $x=1$ and $x=0.5$, respectively.}
		\label{Result8}}
\end{figure}

{To understand better the fractional Fano interference process, we analyze differential conductance as a function of $eV$ for several values of coupling between the QD and lower MBS ($|\lambda_{2}|$), which allows to verify how the fractional feature is modified by decreasing the degree of MBS non-locality~\cite{RAguado}. Fig.~\ref{Result5} shows that, for smaller values of $|\lambda_{2}|$, the fractional lineshape persists with slight changes in amplitude. However, for $|\lambda_{2}| = 10^{-2}E_{0}$ (purple dashed-line) the fractional resonances invert and, for bigger values, vanish. This behavior suggests that the fractional Fano effect appears just for high degrees of MBS non-locality, i.e, $|\lambda_{2}| \leq 10^{-2}E_{0} \ll |\lambda_{1}|$, yielding $\eta = 0.1$.}

{Fig.~\ref{Result6} shows the differential conductance as a function of $eV$ for distinct Fano interference processes ($0<x<1$). When the lead-QD-lead path is dominant ($x=0$), the conductance reaches maximum $e^{2}/h$, indicating that the MBSs are overlapped via Zeeman field ($\varepsilon_{M}(B)\neq0$). As we enhance the direct lead-lead transport, the Fano-like fractional resonance begins to take shape. Such a behavior can be verified for $x \geq 0.15(q_{b} \leq 1.10)$. For higher values of $x$, which describes the predominance of direct lead-lead tunneling ($x \geq 0.75(q_{b} \leq 0.14)$), the fractional resonances becomes more evident, showing the same lineshape, with small deviations in amplitude. These features state that the fractional Fano interference effect takes place when direct lead-lead tunneling process is dominant over those lead-QD-lead, thus indicating that the measurement system (metallic leads and QD) can distort the MBSs local/non-local signatures due to interference phenomena.}

{In order to have an overview about the influence of Fano interference in the Majorana oscillations, in Fig.~\ref{Result7} we analyze the differential conductance for $eV=0$ as a function of Zeeman field, considering several values of $x(q_{b})$. For the highly non-local situation ($|\lambda_{2}|=0$, $\eta = 0$) [Fig.~\ref{Result7}(a)], we verify that as $x$ increases, the amplitude of oscillations are suppressed until total quench for $x=1~(q_{b}=0)$. Thereby, the enhancement of direct lead-lead tunneling process ($x \geq 0.95$) can destroy the oscillatory behavior at zero-bias voltage ($eV=0$), hiding the information about the overlap between MBSs via Zeeman field. The suppression of oscillations amplitude is also verified for $|\lambda_{2}|\neq 0$, even for a high degree of MBSs non-locality ($\eta \sim 10^{-3}$), as depicted in Fig.~\ref{Result7}(b). The main difference is that for the finite $|\lambda_{2}|$ situation, the oscillatory pattern is not completely quenched for $x=1$.}

{Fig.~\ref{Result8} exhibits how the degree of MBSs non-locality affects the Majorana oscillations at zero-bias voltage. The oscillatory behavior is well defined just for higher non-local situations in all the three interference processes considered here [(a)$x=0$, (b)$x=1.0$ and (c)$x=0.5$]. As we decrease the Majorana non-local property (enhancing $|\lambda_{2}|$), the oscillation pattern is totally suppressed due to the MBSs peak splitting, which points out that the MBSs can experience each other. The data indicates that for $|\lambda_{2}|=0.1E_{0}$ (orange solid line), yielding $\eta = 0.32$, the oscillatory pattern is completely absent, which shows that the presence of well-defined oscillations at $eV=0$ in the differential conductance as a function of Zeeman field is a feature of highly non-local MBSs ($\eta \rightarrow 0$). These findings suggest that our device can work as a fine tunneling spectrometer to investigate the non-local MBSs features, once it catches changes in oscillations amplitude appearing in differential conductance at zero-bias, even for small values of $\eta$.}

\subsection{{Degree of MBS non-locality and experimental protocol}}\label{nonlocal}

{In Sec.~\ref{Introduction}, we recall the concept of degree of MBSs non-locality $\eta$ proposed by Prada \textit{et al.}~\cite{RAguado}, who also indicated a protocol to measure it in a QD-TSNW hybrid system. Such a theoretical proposal was followed by its experimental achievement by Deng \textit{et al.}~\cite{Aguadoexp}. We also introduced that $\eta$ is related to a topological quality factor, as stated in Ref.~[\onlinecite{DJClarcke}]. In this subsection, we present that our simplest effective Hamiltonian (spinless carriers, absence of charging effect and additional ABSs) is also able to catch the information of the degree of MBSs non-locality using the same protocol previously proposed~\cite{RAguado,Aguadoexp}. Before presenting our findings, it is worth mentioning that the QD setup in our device is distinct from the original proposal~\cite{RAguado}. Here, the transport is through the QD, placed between metallic leads and side-coupled to the TSNW, while in previous works~\cite{wire2016,PhysRevB.96.125420,RAguado,DJClarcke,Aguadoexp}, the transport is through the QD-TSNW system, placed between metallic and superconducting leads. Furthermore, in such works the QD belongs to the nanowire structure and, therefore is not a separated entity as in our device.
}

\begin{figure}[t]
	\centerline{\includegraphics[width=3.8in,keepaspectratio]{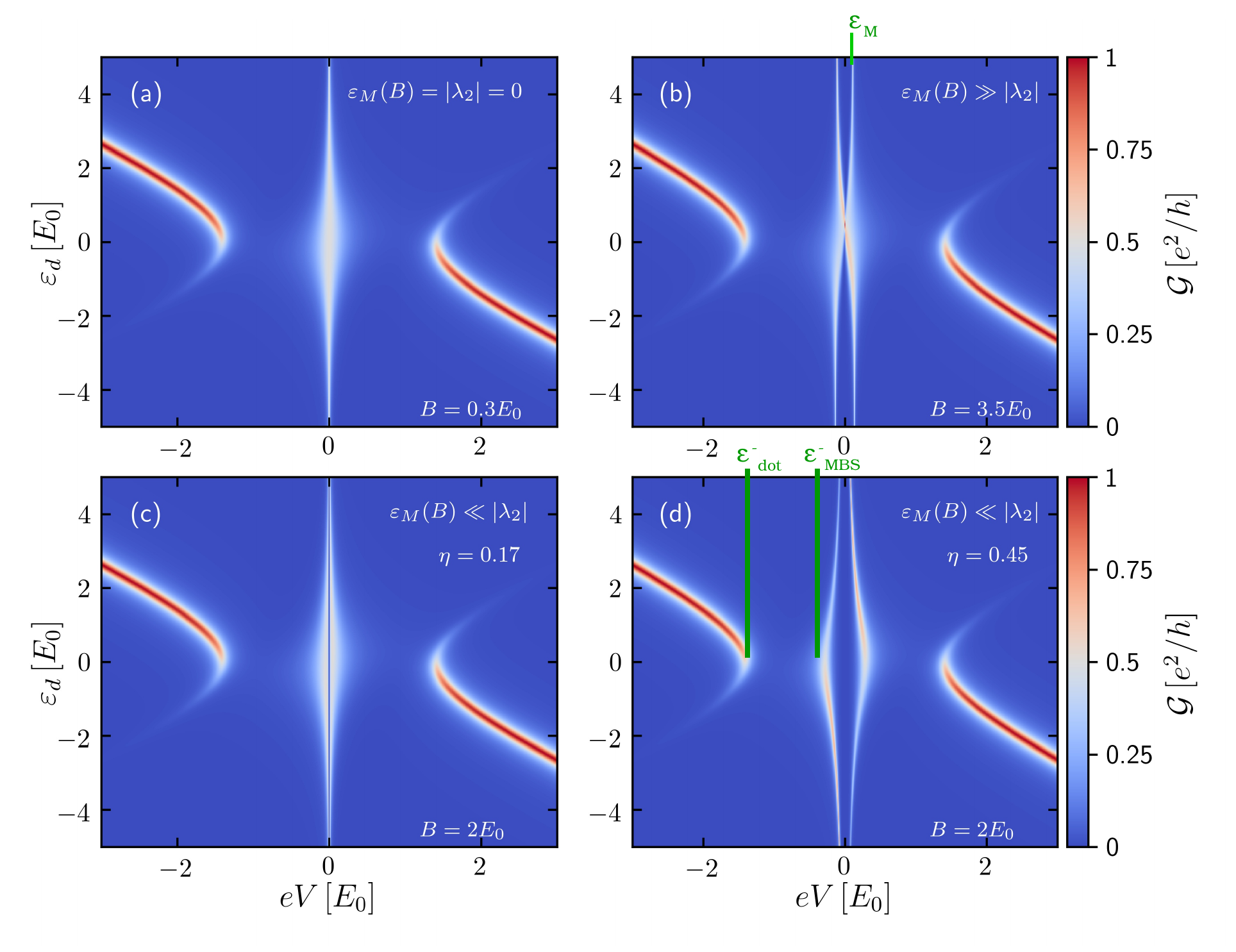}}
	\caption{{Differential conductance [Eq.~(\ref{eq:zbconductance})] as function of both QD energy level $\varepsilon_{d}$ and bias voltage $eV$. Panel (a) shows the highly non-local situation of isolated MBSs ($\varepsilon_{\text{M}}(B)=|\lambda_{2}|=0$). (b) exhibits a non-local situation, wherein the $\varepsilon_{\text{M}}(B)$ is dominant (``bowtie'' overlapping MBSs in Ref.~[\onlinecite{RAguado}]), while (c) and (d) show the case correspondent to ``diamond'' of same reference, described by the dominance of $|\lambda_{2}|$ over  $\varepsilon_{\text{M}}(B)$. The difference between (c) and (d) is the degree of MBSs non-locality $\eta = \sqrt{|\lambda_{2}| / |\lambda_{1}|}$, with $|\lambda_{1}|=1.0E_{0}$ for all the situations considered. The values $\varepsilon_{\text{dot}}^{-}$ and $\varepsilon_{\text{MBS}}^{-}$ allows to obtain experimentally $\eta\approx\Omega$. }
		\label{Result9}}
\end{figure}

{Fig.~\ref{Result9} shows contour plots of differential conductance as a function of bias-voltage $eV$ and QD energy level $\varepsilon_{d}$, for several values of $\varepsilon_{\text{M}}(B)$ and $|\lambda_{2}|$. The QD energy level can be experimentally accessed by a gate potential $V_{d}$, which can be tuned separately from the gate voltage $eV$ between metallic leads by changing both in a compensatory way~\cite{Aguadoexp}. Panel (a) describes the higher non-local situation, \textit{i.e.,} there is no overlap between the MBSs ($\varepsilon_{\text{M}}(B) = |\lambda_{2}|=0$). This highly non-local property is characterized by a plateau at $\mathcal{G}(eV=0)=e^{2}/h$, independent from the value of $\varepsilon_{d}$. It is known from previous works~\cite{wire2016,PhysRevB.96.075161} that Andreev bound states (ABSs) can transmute in a topological MBS as they become merged at zero-energy. However, the ABSs also can coalesce forming near-zero energy mid-gap states in a nontopological regime, mimicking MBS signatures. Such ABSs analysis does not belong to scope of this work, since no additional ABSs were included.}
	
{Now, let us consider the situation in which MBSs overlap with each other via $\varepsilon_{\text{M}}(B) \gg |\lambda_{2}|$. We verify in panel (b) the \textit{``bowtie''} pattern, in qualitative agreement with the same situation reported in Fig.~4(b) of Ref.~[\onlinecite{RAguado}] and 3(b) of Ref.~[\onlinecite{Aguadoexp}]. As indicated in Fig.~\ref{Result9}(b), such a measurement is able to provide the value of $\varepsilon_{\text{M}}(B)$, which is $\approx 0.12E_{0}$ for $B=3.5E_{0}$. Figs.~\ref{Result9}(c)-(d) depict the situation wherein $\varepsilon_{\text{M}}(B) \ll |\lambda_{2}|$, which reveal information about the degree of MBSs non-locality using the same protocol previously stated~\cite{RAguado}: $\Omega\approx\eta$ can be obtained experimentally by the ratio $\varepsilon_{\text{MBS}}^{\pm}/\varepsilon_{\text{dot}}^{\pm}$. Let us pick out the values indicated in Fig.~\ref{Result9}(d): $\varepsilon_{\text{MBS}}^{-} \approx -0.30E_{0}$ and $\varepsilon_{\text{dot}}^{-} \approx -1.5E_{0}$. Since $\Omega^{2} = \varepsilon_{\text{MBS}}^{-} / \varepsilon_{\text{dot}}^{-}$~\cite{RAguado, Aguadoexp}, we find $\Omega \approx 0.45$, in agreement with the theoretical parameters adopted ($|\lambda_{2}|=0.2E_{0}$, $|\lambda_{1}|=1.0E_{0}$ and $\eta = \sqrt{|\lambda_{2}|/|\lambda_{1}|}=0.45$). In such panels, we also confirm the \textit{``diamond''} shape, which was previously verified in Figs.~4(d) and 3(c) of Refs.~[\onlinecite{RAguado}] and ~[\onlinecite{Aguadoexp}], respectively. By comparing panels (c) and (d), it can be noticed that the enhancement of $\Omega$, \textit{i.e.,} the reduction of MBSs non-local properties, is characterized by the opening of the \textit{``diamond''} shape. Qualitative agreement between our results of Fig.~\ref{Result9} and those found in Ref.~[\onlinecite{RAguado}], which were experimentally verified in Ref.~[\onlinecite{Aguadoexp}], evidence that our device can be used to explore the MBSs non-local properties.}

\section{Conclusions}\label{conclusions}

 To summarize, we studied Majorana oscillations in a {T-shaped hybrid device} composed by a QD embedded between a pair of conducting leads and side-coupled to a TSNW hosting zero-energy MBSs at its ends. Analyzing the differential conductance profiles of the system as a function of the applied Zeeman field and bias-voltage $eV$ between the leads, we found that Majorana oscillations are very sensitive to the changes of the regime of Fano interference {and degree of Majorana non-locality $\eta$. This latter can be tuned by changing the coupling between the QD and lower MBS. Unexpected fractional Fano-like resonances were unveiled for high non-local situations ($\eta \rightarrow 0$), in the regime where direct lead-lead tunneling prevails. Moreover, differential conductance as a function of both bias-voltage and energy level of QD revealed \textit{``bowtie''} and \textit{``diamond''} shapes, in qualitative agreement with the original theoretical proposal~\cite{RAguado}, despite differences between the models. Such correspondences indicate that our device also can be used as a tunneling spectrometer to obtain experimentally the degree of Majorana non-locality and investigate its topological properties following the same protocol proposed by Prada \textit{et al.}~\cite{RAguado} and experimentally performed by Deng \textit{et al.}~\cite{Aguadoexp}.}

\section*{Acknowledgments}

We thank the funding Brazilian agencies CNPq (Grant No. 307573/2015-0), CAPES, and S\~ao Paulo Research Foundation (FAPESP) Grant No. 2015/23539-8. I.A.S. acknowledge support from Russian Science Foundation Project 17-12-01359 and Horizon2020 project CoExAN.

\end{document}